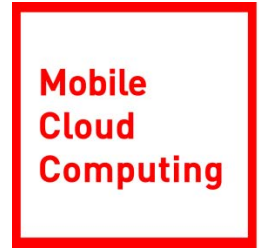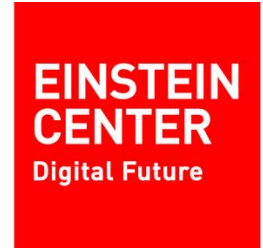

# Technical Report

## No. MCC.2018.1

**Abstract:**

Docker seems to be an attractive solution for cloud database benchmarking as it simplifies the setup process through pre-built images that are portable and simple to maintain. However, the usage of Docker for benchmarking is only valid if there is no effect on measurement results. Existing work has so far only focused on the performance overheads that Docker directly induces for specific applications. In this paper, we have studied indirect effects of dockerization on the results of database benchmarking. Among others, our results clearly show that containerization has a measurable and non-constant influence on measurement results and should, hence, only be used after careful analysis.

# Dockerization Impacts in Database Performance Benchmarking

Martin Grambow, Jonathan Hasenburg, Tobias Pfandzelter, David Bermbach

TU Berlin & Einstein Center Digital Future, Mobile Cloud Computing Research Group

mg,jh,tpz,db@mcc.tu-berlin.de

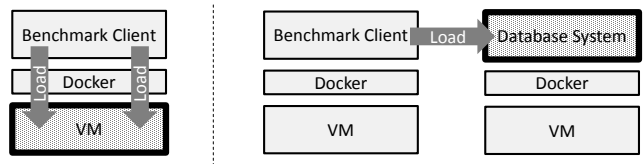

**Figure 1: Different Perspectives in Experimentation with Docker**

## 1 INTRODUCTION

Benchmarking has long been used for the comparison of software and hardware systems or software versions [22]. More recently, it has also been adopted for use cases such as SLA management, service quality improvement, quality control in build processes, service selection, deployment optimization, and others [5]. Benchmarking though, when done right, is surprisingly hard as conflicting goals such as reproducibility, portability, understandability, fairness, ease-of-use, and relevance need to be balanced [4, 20, 23, 33]. When focusing on reproducibility and ease-of-use, an engineer running a systems benchmark is likely to encounter two main challenges: First, correctly installing and configuring both benchmarking client and the system under test (SUT) can be error-prone and challenging, or at least involves a lot of effort. Second, for reproducibility reasons, benchmark runs need to be repeated several times – preferably on a fresh system setup which aggravates the first challenge.

A solution that naturally lends itself to these challenges is to dockerize [27] both benchmarking client and SUT, thus, using containers as a convenient deployment mechanism for preconfigured, ready-to-use experimental setups. This has already been done, e.g., by Palit et al. [29]. However, it is unclear whether this will affect benchmarking results. There are several studies, e.g., [9, 17, 18, 27, 34], measuring the overheads that various applications might incur when running inside containers instead of on bare metal or inside a virtual machine. These, however, all quantify the overhead that is induced by Docker for a certain workload or application. Neither of these studies measures indirect effects of Docker that, for instance, a database benchmark running against an SUT on another machine might experience. For such a benchmark, indirect effects might lead to volatile and unpredictable changes in benchmarking results rendering results at least partially obsolete. See also figure 1 which gives a high-level overview of the different perspectives taken in related work (on the left) and in this paper (on the right).

In this paper, we aim to answer the question whether it is safe to dockerize database benchmarks, i.e., whether dockerization of benchmarking client and/or SUT has observable effects on measurement results. For this purpose, we carefully designed a set of experiments that not only quantifies possible dockerization impacts on benchmarking results but also explores whether different standard settings of both benchmarking client and SUT can further influence potential impacts. For these settings, we used configurations that tend to stress either I/O, compute, or memory. Based on this, as our main contribution, we discuss the results of extensive experimentation with YCSB [11] (a standard cloud storage benchmark) and Apache Cassandra [26] (a widely used NoSQL system) running on Amazon EC2[1]. As a second contribution, we use our observed results to give recommendations and identify implications for database benchmarking with and without Docker.

This paper is structured as follows: In section 2, we start by giving an overview of YCSB, Cassandra, and Docker before discussing related work. Afterwards, in section 3, we describe our experiment design, configuration parameters, and automation tools. Next, in section 4, we discuss the results of our experiments before presenting implications and recommendations in section 5. Finally, we conclude in section 6. This paper is an extended version of [21].

## 2 BACKGROUND AND RELATED WORK

In this section, we give an overview of the software systems used in our experiments – namely, YCSB (section 2.1), Apache Cassandra (section 2.2), and Docker (section 2.3). We also discuss related work in section 2.4.

### 2.1 YCSB

The Yahoo! Cloud Serving Benchmark (YCSB) is an extensible and open-source cloud data storage benchmark tool that was originally developed at Yahoo! Research to facilitate performance evaluations of Yahoo!'s distributed database management system PNUTS [10] and comparisons to other cloud storage solutions [11]. Its extensibility enables the benchmarking of arbitrary database systems. Moreover, it is possible to customize benchmarking workloads by defining various parameters such as record size, the number of operations to be performed, or read-to-write ratio.

---

[1]aws.amazon.com/ec2

YCSB comprises four components: a workload executor that generates the workload, a database interface layer that implements a connection point to communicate with the specified database, a statistics module to record and aggregate benchmarking information, and a component to manage the workload executing threads. YCSB implements the closed workload model [5, 32].

We decided to use YCSB for our benchmarking experiments as it is the de-facto standard for benchmarking of database systems, particularly of non-relational database systems. As such, it provides a collection of connectors for almost all widely used database systems. A second reason for using YCSB was that we wanted to study a "typical" benchmarking scenario. Thus, we had to pick a benchmarking tool that has not only been widely used for experiments with our chosen SUT but one that also works "out of the box". YCSB does just that and has in recent years been widely used with our SUT Cassandra for a variety of purposes, e.g., [1, 6, 11, 24, 25, 30, 31].

## 2.2 Apache Cassandra

Apache Cassandra is a non-relational peer-to-peer database system for massively scalable, distributed data storage and is optimized for performance and scalability rather than strong consistency [26]. Cassandra was originally developed at Facebook for the inbox search problem [26] and combines ideas from Amazon's Dynamo [16] and Google's BigTable [8]. The system is now continued as an open source Apache project and is widely used, e.g., by companies such as Netflix. We decided to use Cassandra due to its popularity and wide adoption in benchmarking as well as in production.

In Cassandra, an update operation is initially stored in an in-memory memtable and concurrently written to the commit log on disk. After a certain amount of time or when the memory limit has been reached, the memtable is flushed to a persistent Sorted Strings Table (SSTable) on the disk. Moreover, Cassandra does not update rows in SSTables but creates new tables with the updated information and a timestamp [14] which have to be merged periodically. This approach is adapted from BigTable [8] and referred to as *compactions* for which Cassandra offers three different strategies: Size Tiered Compaction Strategy (STCS), Leveled Compaction Strategy (LCS) and Time Window Compaction Strategy (TWCS).

Using STCS, a compaction is triggered once a given number of SSTables with a similar size exists. By default, four SSTables are merged into one larger table and these larger tables are merged even further. This strategy is recommended for write-intensive workloads but can spread data items over multiple SSTables which causes slower read operations. In contrast to STCS, LCS is designed for read-intensive applications. LCS initially flushes the data from the memtable into a level 0 SSTable before merging these initial tables into level 1 tables that are about the same size. Starting with level 1, each level consists of disjoint SSTables which link each data item to one SSTable per level and do not spread data items across multiple tables. If the number of tables in a level exceeds a limit, some tables are promoted to the next level on which ten times more tables exist than on the previous one. The non-overlapping tables enable faster reads as there is only one table per level in the worst case. Finally, TWCS is designed for time series data and expiring information. TWCS groups data items into SSTables based on time window information. First, STCS is used to compact all new data of the most recent time window. When a time window ends, all SSTables of this window are merged into a single SSTable and there is no further compaction later on [12, 13].

Cassandra also features a row and key cache with configurable sizes that help to improve read performance for frequently accessed data. The key cache keeps row keys in memory, which speeds up the look-up of rows in SSTables. The row cache, however, stores entire data rows in memory which might be distributed over multiple SSTables. While this can improve read latencies by a large amount for those few data items that are very frequently accessed, it requires much more memory so the row cache is disabled by default [15].

## 2.3 Docker

Docker [27] is a containerization platform developed by Docker Inc. that allows Linux applications, their dependencies, and their settings to be composed into Docker images. These images run as Docker containers on any machine running the Docker daemon, which utilizes kernel namespaces and control groups to isolate running containers and control their set of resources. Docker uses an overlay file system which stores modifications in layers that correspond to a set of differences. New layers are added on top of lower layers and only the topmost layer is writable. Each running container believes to have its own file system, including operation system files, but only one copy of these files is actually present on the Docker host. The default storage driver follows a copy-on-write strategy. If an application inside a running container tries to modify a file, the file system copies the file to the topmost writable layer where it stores the modification. The underlying file is untouched and can be used by other running containers simultaneously, which makes Docker very resource-efficient. Furthermore, Docker also has the option to attach volumes to containers that read and write directly on the host's disk, bypassing the layered file system [3, 27].

Docker simplifies the deployment of applications as an image only has to be created once before it can be deployed on every system running the Docker deamon. Images can be shared via a central platform, which makes it straightforward to download and customize them according to one's own requirements. This makes Docker very attractive for software benchmarking experiments, because an examiner can create multiple Docker images which are based on the same root image but contain different to be benchmarked configurations. Moreover, as images are portable and can be reused, experiments are easy to parallelize and repeat. Finally, it is also possible to limit computational power and resources such as the number of CPU cores or memory used for each running container.

## 2.4 Related Work

There are already several publications trying to quantify the performance overhead of dockerization. For instance, Chung et al. [9] have benchmarked high performance computing applications (HPL and Graph500) running in Docker containers and found remarkable differences to the performance without docker.

In difference to the findings of Chung et al, Di Tommaso et al. [17] have tested Docker's impact on the performance of genome analysis pipelines and concluded that the Docker technology only introduces a negligible performance overhead for their purposes. They executed multiple tests on a cluster of 12 high performance machines and compared the execution time of tasks running in Docker containers to the native performance.

Felter et al. [18] compared the performance of Docker containers and virtual machines utilizing microbenchmarks on a 32 vCPU instance equipped with a not specified but "adequate" amount of memory to execute the given workload. Similar to Di Tommaso et al., they conclude that Docker introduces a negligible computation and memory overhead in most cases, but I/O-intensive workloads should be used carefully as extra cycles are needed for each I/O operation.

Ali et al. [2] benchmarked the performance impact of Docker together with VM technology using microbenchmarks and measured overheads of up to 4%.

In all these publications, the authors measured the *directly* visible overhead of Docker for a certain workload or application. Somewhat comparable to the TLS experiments of Müller et al. [28], we are interested in *indirect* effects of Docker on applications running inside containers. Specifically, we aim to evaluate whether it is safe to dockerize benchmarks, i.e., whether dockerization of benchmarking components does not affect measurement results – neither actual measurement values nor their stability and reproducibility. To the best of our knowledge, this has not been experimentally studied yet. However, such an evaluation is needed as some cloud service benchmarking tools already use Docker as a deployment mechanism. For instance, Palit et al. [29] published a suite of cloud service benchmarking tools as Docker images, Ceesay et al. [7] have built an entire cloud benchmarking framework around Docker, and Ferme et al. [19] have containerized the benchmarking of workflow management systems.

## 3 EXPERIMENT DESIGN

In this section, we describe how we designed our experiments to identify whether dockerization of a benchmarking client and/or an SUT has any observable effects on measurement results. In essence, this implies that we need to compare the results of four different kinds of benchmark runs, as also shown in figure 2: no dockerization at all (I), dockerization of the benchmarking client (II), dockerization of the SUT (III), and dockerization of both systems (IV). We decided to not impose resource limits on running containers for any of our experiments, i.e., each container can freely use all available resources of the VM host. Otherwise, it would have been difficult to compare dockerized and non-dockerized experiments.

By testing different dockerization variants, we can not only determine whether Docker has an impact on benchmarking results but also determine whether the effect is caused at the SUT or the benchmarking client. For each of the four Docker setups, we additionally tweaked various parameters in the benchmarking client and in the SUT to evaluate whether the degree of potential dockerization impact can also be affected by the parameter set. As an infrastructure parameter, we decided to run each experiment not on just one type of machine, so we ran all experiments on both m3.medium and m3.large instances (see table 1) in the AWS region Ireland (eu-west-1b), with client and SUT always using the same instance type but running on different instances within the same availability zone. These instance types are equipped with suitable hardware resources for typical use cases and do not have a credit-based "burst" performance which can absorb temporal load peaks but would also render our results meaningless. In compliance with common practice, each experiment was repeated 5 times.

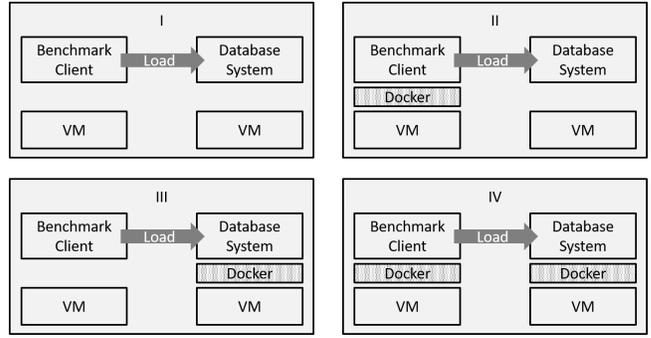

**Figure 2: Dockerization Variants in Database Benchmarking**

| Instance Type | vCPUs | Memory | Storage |
|---|---|---|---|
| m3.medium | 1 | 3.75GB | 4GB SSD |
| m3.large | 2 | 7.50GB | 32GB SSD |

**Table 1: Used AWS Instance Types**

As already pointed out in section 2, we deployed a single Apache Cassandra node as SUT in our experiments. To simulate a "typical" setup, we used default values for all configuration options in our experiments but also changed the compaction strategy and key cache size in later runs to study their effects. We used all compaction strategies presented in section 2.2, namely Size Tiered Compaction Strategy (STCS), Time Window Compaction Strategy (TWCS), and Level Compaction Strategy (LCS). The key cache was either disabled (= 0MB) or set to "auto" (= 4% of the available heap size and up to a maximum of 100MB). We decided to use Apache Cassandra as a single node deployment to reduce the number of influence parameters in our experiments. However, we are well aware that a multi-node deployment is more common. Still, we have seen in past experiments that there are also use cases where a single node Cassandra makes a lot of sense, for instance, as it typically sustains more throughput than MySQL.

On the client machine, we deployed YCSB [11]. We used YCSB's Workload A that consists of 50% reads and 50% writes. For the m3.medium instances, we used 100,000 records and 3,000,000 operations; for the m3.large instances, we used the same number of records but 9,000,000 operations to achieve sufficiently long-running experiments. No other configurations were modified. As seen in preparatory experiments, this results in about 20 minutes per benchmark run for both instance types. In initial test runs, we also experimented with workload B but found results similar to the ones for workload A.

For our experiments, we wanted to test both maximum throughput benchmarks as well as latency benchmarks which are typically run at medium resource utilization. As YCSB controls the generated workload intensity through the thread count (closed workload model), we ran a set of preparatory experiments to identify suitable thread counts. In these experiments, we set Cassandra's compaction strategy to TWCS and the cache size to 0. In these experiments, we noted that the throughput does not increase beyond 50 threads (m3.medium), and beyond 150 threads (m3.large) while both read and write latency continue to rise linearly. Based on these results, we decided to run all m3.medium experiments with 10, 25, and 50 threads, and all m3.large experiments with 30, 75, and 150 threads. Table 2 gives an overview of the parameter sets which we used in our experiments; in total we ran 720 benchmarking experiments.

| Machine | Parameter | Variations |
|---|---|---|
| Client | Docker | Yes, No |
| | Thread Count | 10, 25, 50 OR 30, 75, 150 |
| SUT | Docker | Yes, No |
| | Key Cache Size | 0, Auto |
| | Compaction Strategy | STCS, TWCS, LCS |
| General | Run | 1, 2, 3, 4, 5 |
| | Instance Type | m3.medium, m3.large |
| Total Number of Experiments | | 720 |

**Table 2: Experimental Parameter Variations**

To ensure a consistent environment, we highly automated the benchmarking process and pre-built all Amazon Machine Images (AMI) and Docker containers used throughout the experiments. The Docker containers and AMIs are all based on amazon-linux 2017.09.1 and only contain the software needed to run the experiments. On the SUT machines, Cassandra 3.11.2 in combination with OpenJDK 8 is either installed directly on the system or in Docker, while the client machines use YCSB 0.12.0 which is also either installed directly on the system or in Docker. For Docker, we explicitly decided to use the default configuration options as these are most likely to be used in practice. This means that data was stored in the layered file system of Docker and the network was configured to run in network bridge mode. For each experiment, the following steps were performed in this order utilizing either the AWS command line interface or secure shell:

(1) Create a new client machine
(2) Create a new SUT machine
(3) Start Cassandra on the SUT with the chosen parameters
(4) Start YCSB on the client with the chosen parameters
(5) Collect results from the benchmarking client
(6) Terminate both machines

We used the cartesian product of the individual parameters shown in table 2 as parameter sets, and executed these steps five times for each set. Particularly, we ran a single experiment for each parameter set before starting the next experimental run as we preferred to have time-based fluctuations (caused by AWS) across experiment runs rather than in between parameter sets.

## 4 RESULTS AND DISCUSSION

In this section, we discuss the results of our experiments. First, we present general results that we could derive by aggregating all benchmark runs with the same thread count and instance type (section 4.1). Then, we evaluate the impact of dockerization on benchmarking results for SUT and client parameter variations in more detail by inspecting individual runs rather than averaged values (section 4.2). Finally, we specifically analyze Docker's impact on benchmarking results when running the experiments with different Cassandra cache settings (section 4.3), and different Cassandra compaction strategies (section 4.4).

### 4.1 General Results

A dockerized component introduces an additional layer of complexity into a system. Thus, from the beginning, we expected a decrease in throughput of systems when using Docker containers (variation II and III of figure 2) and the lowest throughput in fully dockerized setups (variation IV). As expected, we found such impacts on measurement results. In summary, our findings show that the dockerization of benchmarking components typically leads to increasing latencies and decreasing throughputs, especially for larger instance types. However, we also observed that in some cases Docker actually increased throughput.

As estimated from preparatory experiments, benchmark runs were designed to take about 20 minutes to complete. Due to the varying number of running YCSB threads and different Cassandra configurations, our actual experiments took between 15 minutes and 28 minutes to complete. We also monitored the resource utilization of client and SUT. In all experiments, the SUT was the limiting factor as its resource utilization was higher (up to a 100% of resources used) than the client's.

Our results indicate a proportional relationship between read and update latency across our four setup types as an increasing read latency is always coupled with an increasing update latency and vice versa. Moreover, the slope of both latencies is similar when plotted in many cases across all setups. Thus, there seems to be no indirect effect that only affects measurement results of either read or write latencies.

For our analysis, we calculated the average read and update latency grouped by instance type, thread count ($th$), and degree of dockerization (I-IV), so each average value was based on 30 experiment runs. Tables 3 and 4 show relative latency changes of these averaged values compared to the baseline across all four setups: Setup I (no dockerization) is the baseline, Setups II and III reflect partial dockerization and IV corresponds to the fully dockerized setup (see again figure 2). Positive values correspond to increasing latency, i.e., a performance reduction.

In general, our results show an increasing read and update latency of operations if components are dockerized. Especially update operations provoke a significant overhead, on average 8.71% for m3.large instances. On the other hand, we also observed read operation performance improvements for experiments with higher thread counts on the client side. Furthermore, we discovered that the dockerization of components on m3.large instances generally results in a larger overhead than the dockerization on m3.medium instances. Finally, we recognized a trend for larger overheads for read and update operations if the system is not used to full capacity.

| | | Setup | | | |
|---|---|---|---|---|---|
| th | op | I | II | III | IV |
| 10 | Read | 0% | 1.42% | 1.84% | 1.21% |
| | Update | 0% | 1.79% | 5.13% | 4.53% |
| 25 | Read | 0% | −0.09% | −0.01% | 0.69% |
| | Update | 0% | 0.50% | 3.06% | 3.73% |
| 50 | Read | 0% | 2.51% | −0.52% | −0.97% |
| | Update | 0% | 2.72% | 2.66% | 2.20% |
| | Read avg | 0% | 1.28% | 0.44% | 0.31% |
| | Update avg | 0% | 1.67% | 3.62% | 3.49% |

**Table 3: Relative Changes of Average Latency on m3.medium Instances for $th = (10, 25, 50)$ Threads and $n = 30$ Experiments**

| | | Setup | | | |
|---|---|---|---|---|---|
| th | op | I | II | III | IV |
| 30 | Read | 0% | 5.53% | 7.02% | 11.74% |
| | Update | 0% | 5.81% | 7.25% | 12.07% |
| 75 | Read | 0% | 6.81% | 2.50% | 8.85% |
| | Update | 0% | 6.73% | 2.43% | 8.70% |
| 150 | Read | 0% | −0.63% | −2.38% | 1.46% |
| | Update | 0% | 2.96% | 1.43% | 5.36% |
| | Read avg | 0% | 3.90% | 2.38% | 7.35% |
| | Update avg | 0% | 5.17% | 3.70% | 8.71% |

**Table 4: Relative Changes of Average Latency on m3.large Instances for $th = (30, 75, 150)$ Threads and $n = 30$ Experiments**

For m3.medium instances, we observed an overhead of up to 5.13% for update operations with 10 running threads on the client machine. Furthermore, we measured an impact of up to 3.73% for 25 threads, and up to 2.72% for 50 threads respectively for the update operation. However, some read operations induce smaller overheads or even performance improvements. We detected relative differences from -0.97% to 2.51%.

Our findings for m3.large instances are similar to the m3.medium results but at a larger scale. For update operations, we measured relative overheads of up to 12.07% with 30 YCSB threads and 5.36% with 150 YCSB threads. As for the medium instances, we also observed some performance improvements for read operations if the SUT is used to full capacity. All in all, the relative impact of Docker on the average latency for read operations varies between -2.38%

and 11.74%.

Already this short analysis of aggregate values indicates that the dockerization of database benchmarks should only be done after a careful analysis and not only for reasons of convenience.

## 4.2 Median Experiment Runs

In the second part of our analysis, we inspected the impact of dockerization on the benchmark for SUT and client parameter variations in more detail. As described in section 3, we repeated each experiment for an individual parameter set five times to account for random fluctuation. As usual (or at least not unusual) when experimenting in the cloud, we found a few outlier measurements. We, therefore, decided to report the results of the median runs instead of calculating averages as the median is more stable in the presence of outliers. In the following, the median run for a specific, individual parameter set is defined as the run with the median throughput, as throughput and latencies are interconnected. Again, each set of five experiments was not executed in direct sequence to lower the impact of temporal performance differences of AWS on the measurements. For experiments with higher variance, we report the results of all experiment repetitions. In this context, please, note that YCSB reports the average latency in each test repetition. We, here, report the median of five such average latency values.

The experiments plotted in figure 3 show the results for m3.large and m3.medium instances, the plotted Cassandra compaction strategy is TWCS and the cache was disabled. YCSB ran with 50 threads on the m3.medium instance and 150 threads on the m3.large instance. Almost all experiments have very similar curves so we refrain from showing these (almost identical) figures for reasons of legibility; the full data set is available as open source[2]. Dockerizing any component, either the YCSB client or the Cassandra server, typically increases the average latency for read and update operations slightly. Dockerizing the respective other system as well further increases the average latencies. The standard deviation of latencies (across the respective five experiment repetitions) varied from 0.21 ms to 1.38 ms for m3.medium instances, and from 0.30 ms to 0.95 ms for m3.large instances respectively.

We found very few experiments where results deviated from the patterns set in figure 3. We show the two most exceptional results in figures 4 and 5; here we plotted all five benchmark runs. In the corresponding experiments, Cassandra ran without caching and used STCS compaction. Moreover, there were 150, respectively 50 running YCSB threads depending on the AWS instance type. In figure 4 (m3.large), we observed the highest latencies in the non-dockerized setup, but also increasing latencies for setup (III) and (IV) compared to (II). Furthermore, we determined a standard deviations of 1.65 ms for update and 4.61 ms for read operations in the experiment runs for the non-dockerized setup (I) while the lowest latency of an individual run was around 18.5 ms, which is a lot lower than the median of 21 ms. Figure 5 (m3.medium) also shows decreasing latencies if the SUT is dockerized. However, the standard deviation in these runs is below 0.95 ms for setup II and around 2 ms for the other setups.

---

[2]https://github.com/martingrambow/dockerExperiments

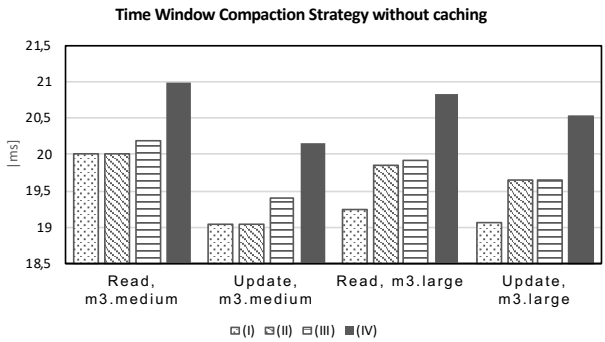

**Figure 3: Typical Result: Dockerization Increases Latency (stddev: 0.21 to 1.38ms for m3.medium; 0.3 to 0.95ms for m3.large)**

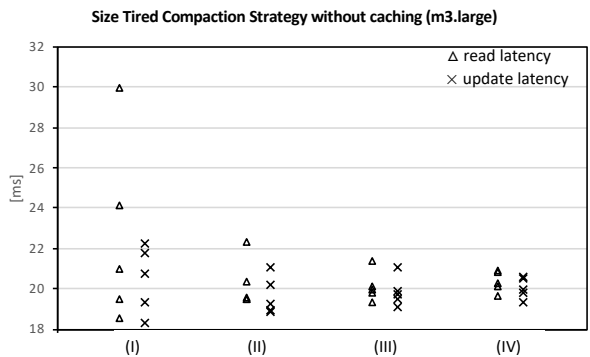

**Figure 4: Atypical Result: Increased Variance of the Average Latency in Non-Dockerized Setups on m3.large Instances**

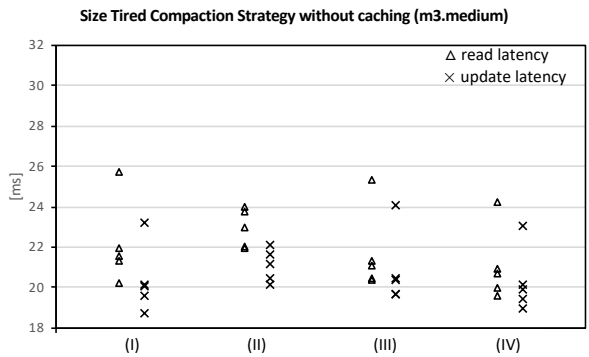

**Figure 5: Atypical Result: Increased Variance of the Average Latency in Non-Dockerized Setups on m3.medium Instances**

Overall, we believe that the more unexpected results might be caused by random performance variation of the underlying VMs. However, if this were a result of dockerization instead (in some scenarios dockerization leads to performance increases), then this would further emphasize our point that dockerization of benchmarking experiments should be considered carefully.

## 4.3 Cache Settings in Cassandra

Besides the general influence of Docker on the benchmarking process, we also evaluated different settings in the configuration of Cassandra. First, we evaluated two different settings for the partition key cache parameter of Cassandra. We either disabled the cache or used the automatic key cache mechanism of Cassandra which aims to speed up read requests. Using caches stresses the RAM of the SUT more than a setup without caching.

Our experiments indicate that changing the key cache setting does neither influence the performance of Cassandra itself (for our experiment workload), nor does it influence the dockerization effects we reported in section 4.2.

Especially for light workloads, our results clearly show that caching neither has an effect on the performance of Cassandra nor does it result in an additional influence on the dockerization effects: Neither the YCSB client nor Cassandra showed extensively changed relative latencies compared to the other setups. However, we could not clearly show this for heavier workloads (number of threads = 50, resp. 150) as the overall variance of results (no matter whether dockerized or not) was too high to draw a well-founded conclusion. We assume that this increased variance is caused by the high utilization of the virtual machines which might lead to conflicts in the scheduling of threads and thus results in unstable latencies, especially in the presence of noisy neighbors. A more detailed analysis for heavy workloads requires significantly more experiments and further investigation; this is beyond the scope of our paper.

## 4.4 Compaction Strategies in Cassandra

Besides the cache setting, we also evaluated how Docker influences the benchmarking results when running the experiments with different Cassandra compaction strategies. All strategies cause a temporal disk usage peak while running since the compactions only run at specific times and for a short period.

On m3.medium instances, we found that the chosen compaction strategy has a negligible impact on Cassandra when running a light workload. This, of course, implies that the compaction strategy does not have an influence on the existing dockerization effects. On m3.large instances, again lightly loaded, we found that the compaction strategies have an influence on Cassandra (with leveled compaction being the fastest) but this influence is constant across different dockerization setups. This means that the compaction strategy does not cause an additional effect on the dockerization impacts.

On both machine types, we again found large overall performance variability across experiment runs (independent of the degree of dockerization). Again, we believe that the high utilization of the underlying VM leads to scheduling conflicts within the machine and is, thus, particularly vulnerable to noisy neighbors leading to unstable performance. It appears that the performance variance of the underlying cloud infrastructure exceeds potential dockerization impacts.

## 5 IMPLICATIONS

As described in chapter 4, our results clearly show an influence of Docker on the results of database benchmarking experiments. However, this influence is not constant as it varies for different configurations and can be up to 12% (in our experiments) which may be still acceptable for some use cases. So what does this mean for database benchmarking?

First, results of dockerized benchmarks can be acceptable when comparing different database systems. In such a case, the absolute measurement values should be disregarded; the ordering of system alternatives, however, is unlikely to change if the difference between alternatives is sufficiently large – e.g., greater than 20-30%.

Second, benchmark setups should be as close as possible to the production environment that they try to emulate [5]. This, however, implies that when production systems are supposed to be dockerized, benchmarking systems also need to be dockerized when measurement accuracy matters.

Third, when evaluating system configurations or implementation alternatives, it may be an option to dockerize the benchmark (as is commonly done in build processes). However, such results can only be used to achieve a general "feeling" of a system's performance. Actual numbers are too unreliable.

Fourth, in many cases it may be acceptable to dockerize the benchmark as long as it stays dockerized and no configuration changes are made. Absolute values should still not be compared to non-dockerized values directly (or should be taken with a grain of salt), but the workload generation and measurements should be stable enough for comparison over multiple measurements.

Finally, repeating sufficiently long experiments is always important in benchmarking. When using dockerization, however, even more repetitions and longer experiments should be used to identify random fluctuations introduced through another layer of indirection.

## 6 CONCLUSION

In this paper, we acknowledged the growing importance of Docker for the benchmarking community. However, we noted that the effect of dockerization on benchmarking results is still unclear as prior studies have only measured the direct overhead of dockerization on certain workloads or applications. Therefore, we extend prior attempts to quantify the influence of Docker by also measuring indirect effects that, for instance, a database benchmark running against an SUT on another machine might experience.

We deployed the state-of-the-art database system Cassandra and the widely used database benchmarking tool YCSB on one machine each to perform a series of 720 experiments with and without Docker in order to better understand how certain small changes to either of these systems, such as disabling caches or an increased thread count in YCSB, affect the overall performance and database request latency as reported by YCSB. We particularly chose a standard combination of benchmark and SUT which is frequently used in research and practice to study how an out-of-the-box solution might be affected by dockerization.

Our results show that the dockerization of benchmarking tool and/or SUT indeed has an influence on benchmarking results as we observed latency increases of up to 12% and decreases of up to 2%. Furthermore, we observed that the impact of Docker on latency is less prominent for heavy workloads.

Based on our results, we derived a number of implications regarding our main question whether it is safe to dockerize database benchmarks: Overall, dockerization of benchmarking system and SUT is not a good idea. Dockerized results can and should only be used for a general ranking of SUTs as the concrete influence of Docker on measurement results depends on too many factors. However, this also implies that if the production environment is dockerized, then the corresponding benchmarking environment should be as well.

In future work we would like to extend our research in a number of different directions. First, as Cassandra is usually deployed in a distributed fashion, we would like to study Docker's impact on a distributed Cassandra cluster that consists of multiple different machines which we decided not to do in this paper to reduce the parameter space. Furthermore, we would like to run an additional set of experiments to better understand the impact of Docker's layered file system on benchmarking results. For that, we would like to measure how the results of the experiments presented in this paper change if Cassandra stores its data inside a volume that is accessible from outside of the container rather than storing data inside the container's layered file system. Finally, considering all of our findings and drawn implications, we would like to build a framework for benchmark automation that makes careful use of dockerization based on a knowledge base of dockerization impacts.